\documentclass[aip,jcp,amsmath,amssymb,reprint]{revtex4-1}

\usepackage{graphicx}
\usepackage{dcolumn}
\usepackage{bm}
\begin{document}

\title{Density functional study of weak ferromagnetism in a thick BiCrO$_3$ film }
\affiliation{Department of Physics, University of Texas, Austin, Texas 78712-0264}
\author{Jun Ding} 
\affiliation{Institute of Physics and Beijing National Laboratory for condensed Matter Physics, Chinese Academy of Sciences, P. O. Box 603, Beijing 100190, People's Republic of China}
\author{Yugui Yao}
\affiliation{Department of Physics, University of Texas, Austin, Texas 78712-0264}
\affiliation{Institute of Physics and Beijing National Laboratory for condensed Matter Physics, Chinese Academy of Sciences, P. O. Box 603, Beijing 100190, People's Republic of China}
\author{Leonard Kleinman}
\affiliation{Department of Physics, University of Texas, Austin, Texas 78712-0264}
\date{\today}

\begin{abstract}
Very thick films of BiCrO$_3$ have been grown on a SrTiO$_3$ substrate, maintaining a tetragonal lattice up to thicknesses of 14,000\AA. Assuming we can treat films of this thickness as bulk crystals, we first calculated the experimentally undetermined atomic positions within the unit cell with the measured lattice constant of the film, then relaxed the lattice constants along with the atomic positions. All the calculations result in an antiferroelectric crystal with the {\it Pnma} space group with noncollinear Bi displacements. We find G-type antiferromagnetism with a spin-orbit induced weak ferromagnetic component, however, the weak ferromagnetic component can cancel if the antiferromagnetic spins are oriented along a particular one of the three 2-fold rotation axes.
\end{abstract}


\maketitle

\section{Introduction}
Ferroelectric antiferromagnetic crystals often have a weak spin-orbit induced ferromagnetic component. Fennie \cite{Fennie2008} has recently shown that for a particular class of these materials, a reversal of the ferroelectric polarization results in a reversal of the ferromagnetic polarization. This gives these materials potential device applications as well as scientific interest. Unfortunately, the ferromagnetic moment is probably too small to be of practical use and the Currie temperatures are usually well below room temperature \cite{Varga2009}. Dzyaloshinsky \cite{Dzyaloshinsky} postulated that weak ferromagnetism could occur only if the ferromagnetic state maintained the magnetic (color) group symmetry of the antiferromagnetic state.  For a crystal with a trigonal axis, he found that ${\bf D}\cdot{\bf S}_1\times{\bf S}_2$ was an allowed term in the Hamiltonian which favored spin canting, where {\bf S}$_1$ and {\bf S}$_2$ are neighboring spins and {\bf D} is a vector along the single trigonal axis. Moriya\cite{Moriya} then found a formula for {\bf D}. We will show how the Dzyaloshinsky-Moria (D-M) theory manifests itself in a crystal with three 2-fold symmetry axes, {\it i.e.} how is the 2-fold-axis along which the D vector points to determined? Does that axis depend upon the orientation of the spins prior to the inclusion of the spin-orbit coupling(SOC)? 

     BiCrO$_3$(BCO) was first synthesized over forty years ago by Sugawara {\it et al.} \cite{Sugawara1968} Their finding of a triclinic low temperature crystal structure is, however, in conflict with some, but not all, more recent monoclinic determinations. There is a structural phase transition at about 430 K, above which it agreed \cite{Sugawara1968,Niitaka2004,Darie2008,Belik2008} that BCO crystallizes in an orthorhombic structure with the {\it Pnma} space group (no.62). Below 430 K x-ray powder diffraction patterns (XPD) were interpreted \cite{Niitaka2004} to yield a $C2$(space group no.5) monoclinic structure. The lack of inversion symmetry implies a ferroelectric crystal. While, on the other hand, two different neutron powder diffraction (NPD) \cite{Darie2008,Belik2008} experiments led to the conclusion that, although the lattice parameters were almost identical to XPD data \cite{Niitaka2004}, the space group is the inversion symmetric $C2/c$(no.15). This coupled with the observed positions of the Bi atoms lead to the conclusion that BCO is antiferroelectric. G type antiferromagnetism is found \cite{Darie2008,Belik2007} to occur at about 110 K with a parasitic ferromagnetic moment whose remnant value is 0.025 $\mu_B$/Cr. The magnetic moment on each Cr atom was determined \cite{Darie2008} to be about 2.6 $\mu_B$ and to point along the {\it b} two-fold axis but to rotate by 50$^o$ in the {\it bc} plane as the temperature is lowered between 80 K and 60 K.

    Kim {\it et al.} \cite{Kim2006} (KLVC) and Murakami {\it et al.} \cite{Murakami2006} (MFLL) have grown films of BCO with different results. KLVC grew films between 200 and 1400 nm thick on SrTiO$_3$ (STO) substrates with 8 nm thick epitaxial SrRuO$_3$ layers as bottom electrodes. Their x-ray diffraction studies revealed a tetragonal lattice with  $a_0 =  c_0 = 3.888${\AA} and $b_0 = 3.902 $\AA. They find the Curie temperature T$_c$ = 140 K  and a weak  ferromagnetic moment (FM) of 0.012 $\mu$$_B$/(f.u.) obtained from a sample cooled down to a very low temperature in a 2 kOe in-plane field. This weak FM can be compared with bulk crystal value \cite{Belik2007} of 0.025 $\mu$$_B$/(f.u.) obtained from a hysteresis curve. They also measured {\it P(E)}, the electric polarization at 15 K as a function of a 1 kHz applied electric field in the direction normal to the film, and obtained the doublely looped hysteresis curve characteristic of antiferroelectrics. At E = $\pm$700 kV/cm, where the {\it P} hysteresis loops closed, {\it P} had the rather small value of $\pm$12$\mu$C/cm$^2$. The magnetic and electric polarization data reported by KLVC were on films 750 nm thick; which corresponds to  a thickness of 1922 single f.u. unit cells. MFLL grew BCO films on LaAlO$_3$ (also on STO and NdGaO$_3$ but for which no data was reported). They obtained a triclinic structure with $a_0=c_0= 3.9$\AA, $b_0 = 3.88$\AA, $\alpha = \gamma =90.6^o$, and $\beta = 89.1^o$ which is only a small distortion away from being tetragonal and in fairly good agreement with the original \cite{Sugawara1968} room temperature ground state measurement. They obtained a weak FM of 0.05  $\mu$$_B$/(f.u.) from their strongly hysteric curve. Using piezoelectric force microscopy MFLL find their film to be ferroelectric. This probably is consistent with their triclinic structure but they could also be observing a surface effect. Except for a TEM image of a 200 nm BCO film, they did not report any film thicknesses. Recently, David {\it et al.} \cite{David} reported that, using X-ray diffraction and transmission electron diffraction at 300 K, they found three phases in BCO films grown on SRO. The two they were able to identify were $C2/c$ and {\it Pnma}. This is the first independent experimental confirmation that {\it Pnma} BCO can exist below 430 K.

    There exist four calculations for BCO of which we are aware, none of which contained the SOC necessary to obtain the parasitic ferromagnetism. Hill {\it et al.} \cite{Hill2002} did the calculation for a hypothetical cubic perovskite structure. They found a soft phonon mode from which they concluded that BCO should be an A type antiferrodistortive or antiferroelectric, {\it i.e.} ferroelectric planes with antiferroelectric coupling between planes. They also predicted it was a G type antiferromagnet, {\it i.e.} every Cr atom surrounded by six Cr atoms with opposite spin. Xu {\it et al.} \cite{Xu2009}  again found G type antiferromagnetism assuming the $C2/c$ structure and calculated a N$\acute{e}$el temperature more than two times larger than the experimental value. Baettig {\it et al.} \cite{Baettig2005} calculated the ferroelectric polarization and exchange coupling of BCO in the BiFeO$_3$ rhombohedral structure which has not yet been observed in BCO. Ray {\it et al.} \cite{Ray2008} calculated stress in the prototypical cubic structure for different magnetic structures and studied the effects of magnetic ordering on the phonon frequencies in cubic perovskite structure.

    Our goal in this paper is to confirm the structure of a thick epitaxial film and to determine its properties. We do this by keeping the lattice constants fixed at those of the epitaxial film and calculating the positions of the atoms in the unit cell, finding the same {\it Pnma} structure as that when we allow the lattice constants to relax to their equilibrium values, and the same as the experimental high temperature phase. The equilibrium and epitaxial crystals are found to be antiferroelectric and antiferromagnetic, but have spin-orbit induced spin tilting. the tilting direction depending upon a 2-fold axis along which the antiferromagnetic spins are originally aligned, but the direction of the D-M {\bf D} vector is found to be independent of the original alignment. 

\section{Computational Details}
Because the KLVC magnetic and electric measurements were conducted on films  1922 single f.u. unit cells thick, we feel confident in treating these films as bulk crystals. Calculations on the film were performed using the measured lattice constants, while determining the experimentally undetermined atomic positions within the unit cell.(By "film" we always mean relaxed atomic positions but unrelaxed experimental lattice constants.) The calculation was then repeated with both the lattice constants and atomic positions relaxed simultaneously to determine the {\it Pnma} ground state. For both calculations simulated annealing with no symmetry constraints was applied.

    These calculations were performed using the projected augmented wave method \cite{Blochl1994} as implemented in the VASP code \cite{Kresse1996}. For consistency with the International Tables \cite{Boston1985}, we take {\it a} and {\it c} to lie in-plane, while {\it b} is normal to it. The four f.u. film lattice constants are  $a = c = \sqrt{2} c_0$ and $b = 2 b_0$ with {\it a} and {\it c} rotated by 45$^o$. The local density approximation (LDA) for exchange and correlation(xc) was treated with the Perdew-Zunger parameterization \cite{Perdew1981} with and without a Hubbard \cite{Dudarev1998} {\it U}-{\it J} of 2.2 eV. The generalized gradient approximation (GGA) in the Perdew-Wang 91 form \cite{Perdew1992}, also with and without {\it U-J} = 2.2 eV was used as well. The outer core Bi 5{\it d} and Cr 3{\it p} electrons were treated as valence electrons. The energies with the SOC included were calculated self consistently using the atomic positions and lattice constants determined without the SOC. The atomic positions were iterated on until the largest force on any atom was 0.005 eV/\AA, resulting in a 10$^{-6}$ eV convergence of the total energy. (All convergence tests were done for the GGA case.) All energies and lattice constants are per four f.u. (20 atom) unit cell. We used a $7\times4\times7$ Brillouin zone sampling throughout except in one case where a $10\times6\times10$ sampling revealed a convergence of 0.02 meV. A 500 eV cutoff in the plane wave expansion was used which resulted in the rather poor convergence of 70 meV in the total energy when compared with a 700 eV cutoff but a satisfactory convergence of 1 meV (2\%) in {\it $\Delta$E}, the bulk crystal energy minus the film energy.

    We do not discuss the bulk low temperature ground state because whether it is triclinic or monoclinic seems to depend on impurity content, imperfections, and domain formation, which is beyond the scope of this paper.

\section{Results and Discussions}

\begin{table}
\begin{ruledtabular}
\caption{$\Delta$E, the fully relaxed bulk crystal energy minus the film energy, in eV, calculated with four xc functionals.}
\begin{tabular}{cccc}
         &     $\Delta$E(No SOC) &   $\Delta$E(SOC)   \\
\hline
GGA    &  -0.04864  &  -0.03966   \\
GGA+$U$  &  -0.12025  &  -0.11986   \\
LDA    &  -0.52597  &  -0.50464   \\
LDA+$U$  &  -0.32089  &  -0.30162  \\
\end{tabular}
\end{ruledtabular}
\end{table}

In table I we display {\it $\Delta$E}, calculated using the GGA, GGA+{\it U}, LDA, and LDA+{\it U} xc energy functionals. The GGA yielded the smallest energy difference, 40 meV (49 meV without SOC), which is quite small for a 20 atom unit cell. Although small, this might be enough to destabilize a 1400 nm thick film, so, probably, actual value should be even smaller since the strained film has been observed. FIG.1. is the GGA film density of states, showing an energy gap of about 1 eV. All xc functionals resulted in semiconductors for both the film and relaxed crystal. The lowest peak which contains 24 electrons per 4 f.u. unit cell is almost pure oxygen {\it s}.  The  peak at -11 eV, containing 8 electrons, mainly comes from Bi {\it s}. The isolated peak at the top of the valence band contains 12 electrons. They are mainly Cr {\it d } electrons, which have mixed t$_{2g}$ and e$_g$ symmetry, and negligible O {\it p} ones. The 72 electron band below it contains O {\it p} hybridized with Cr {\it s, p,} and {\it d} and with Bi {\it s} and {\it p}. The DOS includes SOC but the projections were obtained without it. 

\begin{figure}[h]
\begin{center}
\scalebox{0.46}{\includegraphics {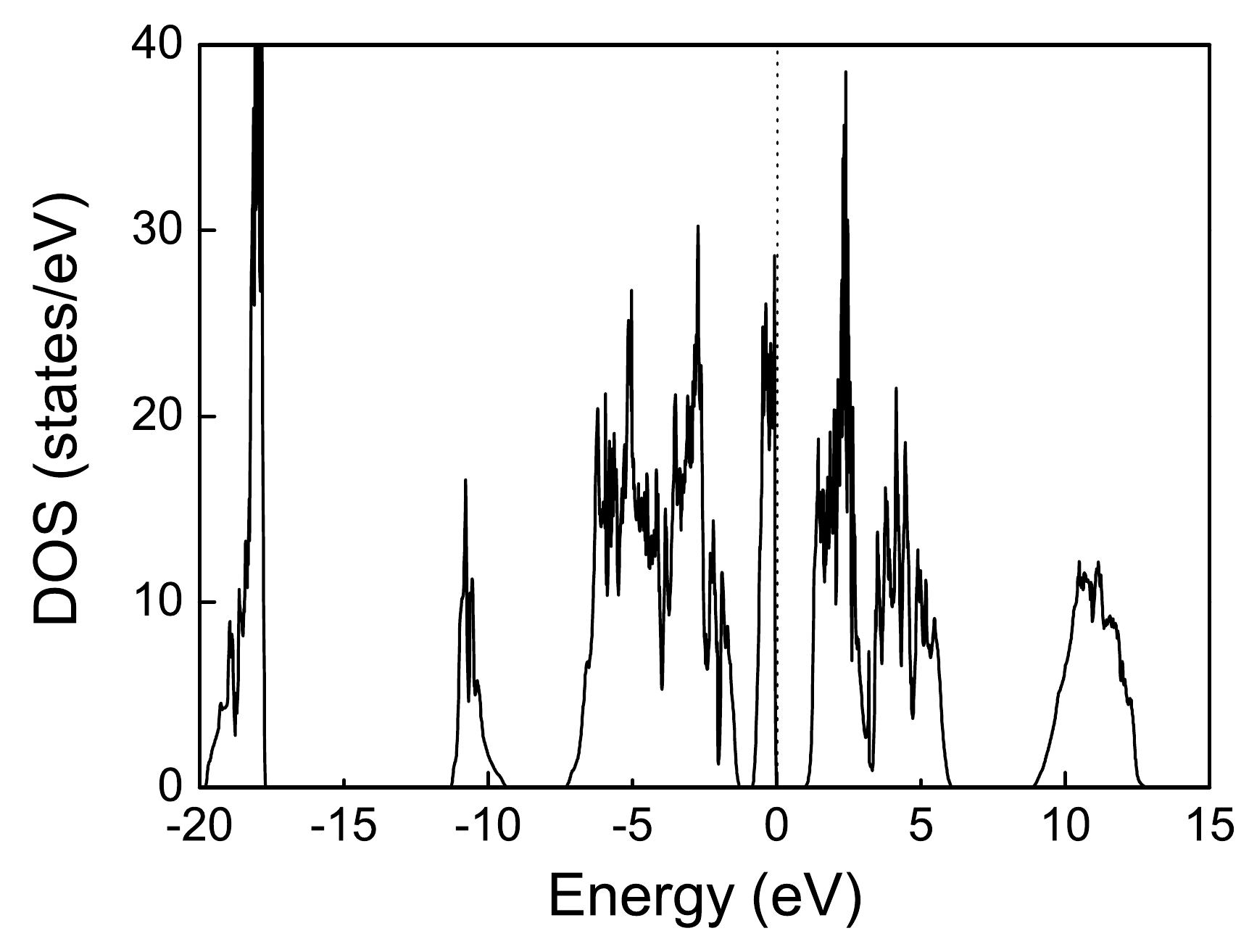}}
\caption{ Total GGA DOS(with spin orbit coupling).}
\end{center}
\end{figure}

\begin{table}
\begin{ruledtabular}
\caption{High temperature experimental bulk \cite{Darie2008} and experimental film \cite{Kim2006} lattice constants (in \AA) and volumes (in \AA$^3$), compared with four calculated results.}
\begin{tabular}{ccccc}
      &  a     &  b    & c    & Volume  \\
\hline
Bulk  & 5.5427 & 7.7524 & 5.4255 & 233.130   \\
Film  & 5.498  & 7.804  & 5.498  & 235.939  \\
GGA   & 5.612  & 7.787  & 5.442  & 237.783  \\
GGA+$U$ & 5.645  & 7.816  & 5.461  & 240.931  \\
LDA   & 5.455  & 7.612  & 5.344  & 221.880  \\
LDA+$U$	& 5.497  & 7.647  & 5.357  & 225.194  \\
\end{tabular}
\end{ruledtabular}
\end{table} 

    In table II we list the experimental bulk crystal, the film, and the calculated lattice constants, as well as the unit cell volumes. The experimental bulk lattice constants \cite{Darie2008} were determined at 470K where the ground state is orthorhombic with {\it Pnma} symmetry and differ slightly from those at 500K \cite{Niitaka2004} and 490K \cite{Belik2008}. Note that {\it a} = {\it c} in the film, but not when the lattice constants are relaxed, either theoretically or experimentally. The GGA relaxed volume however, is only 0.78\% larger than the film's volume. The relaxed bulk crystal volume is larger than the bulk experimental volume by 2\%, but if one tries to factor in what the experimental value would be if the bulk {\it Pnma} crystal structure were stable down to 0 K, the error is more like 4\% or 5\%. This is typical for density functional calculations where errors of 2\% in lattice constants are common. Note that with an appropriate average of the GGA and LDA potentials, we could fit either the film or the experimental bulk volumes, get closer to the experimental lattice constants, and, thus, reduce $\Delta$E and improve the stability of the film.

\begin{figure}[h]
\begin{center}
\scalebox{0.4}{\includegraphics {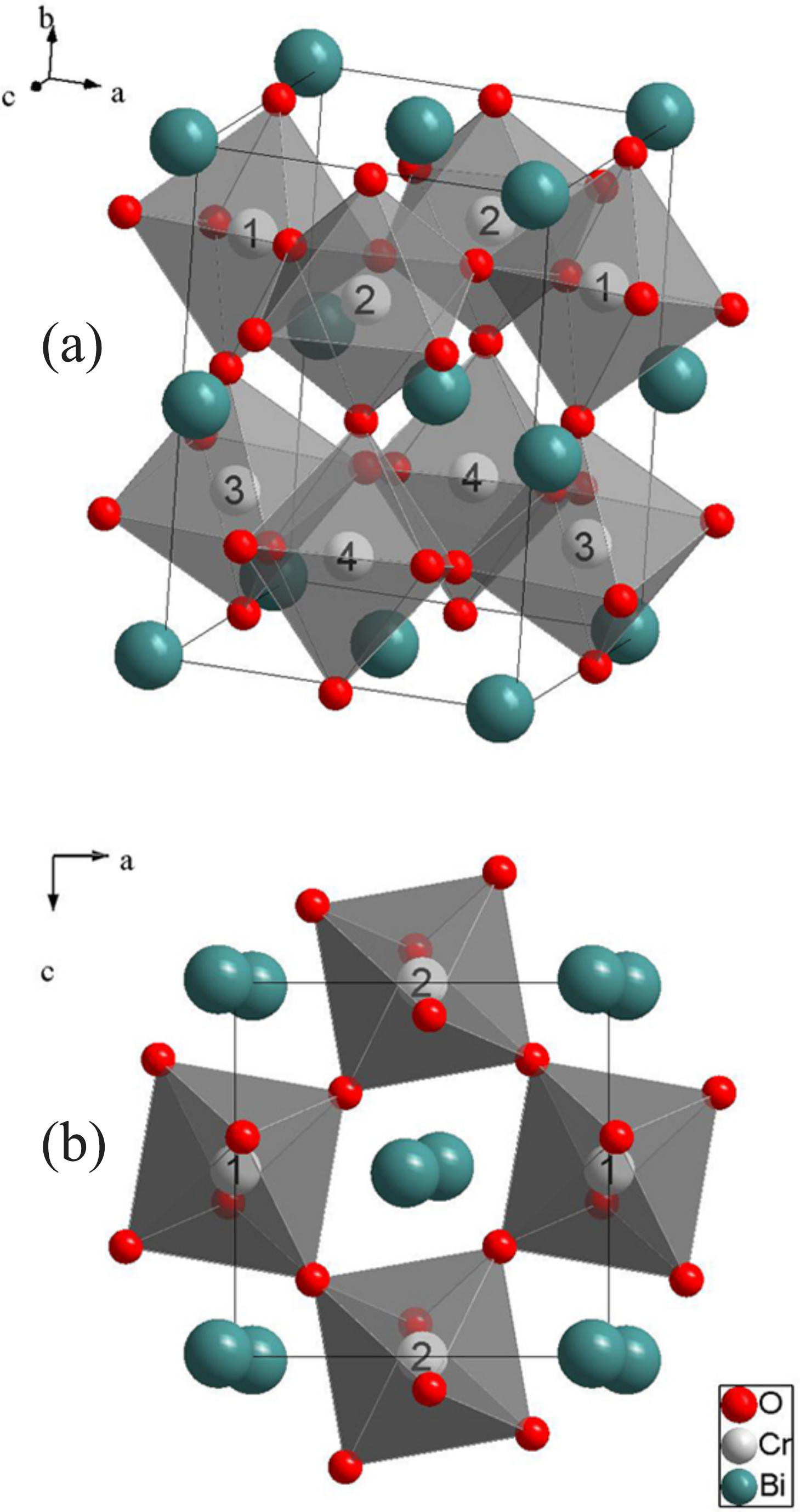}}
\caption{ (color online). Calculated structure of the BCO film showing the tilt of the O octahedra and the Bi displacements. In (a) one can see the displacement of the middle Bi plane with respect to the top and bottom Bi Planes. In (b) the same displacements are seen from  a top view. Note this lattice has been translated by -b/4 relative to the coordinates used in table III.}
\end{center}
\end{figure}

\begin{table}
\begin{ruledtabular}
\caption{GGA calculated film and fully relaxed structure parameters together with experimental high temperature data \cite{Darie2008}. See Ref.19 for Wyckoff notation and other atomic positions.}
\begin{tabular}{cccccc}
            & atom & Wyckoff& $a$      & $b$      &$c$  \\
\hline
Experiment  & O1   & 8d     & 0.2948  & 0.0390  &0.7015   \\
            & O2   & 4c     & 0.4793  & 1/4     &0.0811  \\
            & Bi   & 4c     & 0.0432  & 1/4     &-0.0051  \\
            & Cr   & 4b     & 1/2     &  1/2    & 0 \\
Film(GGA)   & O1   & 8d     & 0.29356 & 0.04390 &0.70571   \\
            & O2   & 4c     & 0.48092 & 1/4     &0.08687  \\
            & Bi   & 4c     & 0.04576 & 1/4     &-0.00874  \\
            & Cr   & 4b     & 1/2     &  1/2    & 0 \\
Fully relaxed & O1 & 8d   & 0.29546   & 0.04291 &0.70153   \\
 (GGA)        & O2 & 4c   & 0.47696   & 1/4     &0.08755  \\
              & Bi & 4c   & 0.05141   & 1/4     &-0.00927  \\
              & Cr & 4b   & 1/2       &  1/2    & 0 \\            
\end{tabular}
\end{ruledtabular}
\end{table}   

    FIG.2. displays the positions of the Bi atoms and the tilting of the oxygen octahedra around the Cr atoms in the film.  Table III compares NPD measurements \cite{Darie2008} of the atomic positions with GGA calculations for the film and bulk crystal cases. Other xc functionals yield results similar to GGA. First note that, in spite of the fact that the film has a tetragonal lattice, both cases have the same {\it Pnma} space group as the high temperature experimental phase. The Cr atoms are located at inversion centers. One can see that the Bi within a plane are displaced from their positions in an ideal perovskite lattice (in the GGA film, with similar results for other cases) by (0.04576{\bf a} $\pm$ 0.00874{\bf c}) in one plane and by - (0.04576{\bf a} $\pm$ 0.00874{\bf c}) in the other, {\it i.e.} the interplanar coupling is antiferroelectric but the intraplanar {\it a} component is ferroelectric while the {\it c} component is antiferroelectric. Thus BiCrO$_3$ is a mixed A and G type antiferroelectric.  The Bi displacement together with the rotation of the oxygen octahedra are the origin of the antiferroelectroic nature of {\it Pnma} BCO.

\begin{table}
\begin{ruledtabular}
\caption{Energy (in meV) of BCO film for different spin orientations relative to the {\it b} axis, {\it i.e.} to the (010) direction,calculated in the GGA and GGA+$U$.}
\begin{tabular}{ccc}
oriention &  GGA  &    GGA+$U$   \\
\hline
010    &  0       &  0       \\
100    &  0.162   &  0.039   \\
001    &  0.160   &  0.003   \\
101    &  0.153   &  0.006   \\
011    &  0.077   &  0.005   \\
110    &  0.072   &  0.004   \\
\end{tabular}
\end{ruledtabular}
\end{table}

    We found that for all four xc functionals, the magnetic ground state is, as expected, G type. We found the C type, A type, and ferromagnetic states are 35.93, 153.76, 208.84 meV above the G type, respectively. These results are GGA without the spin-orbit interaction. The SOC was introduced with all the Cr spins initially aligned in one of several directions, some of whose energy relative to the {\bf b} alignment are displayed in table IV.  Although the SOC energy is 2.02 eV, the anisotropy energy differences are small and the spins remained in their original orientations, except for the spin-orbit induced deviations displayed in table V. When the SOC was turned off but noncollinearity still allowed, the spins returned to their original collinear antiferromagnetic orientations. The magnetic moment of 2.64 $\mu$$_B$ on each Cr agrees well with the value of 2.6 $\mu$$_B$ deduced \cite{Darie2008} from NPD data, albeit for a different crystal structure. We note that for the {\bf b} orientation, {\it a} and {\it c} components are induced on each Cr atom but the {\it a} components are antiferromagnetic and therefore cancel, leaving a SOC induced ferromagnetism of 0.073 $\mu$$_B$ per Cr in the {\bf -c} direction. Similarly, the {\bf c} orientation has 0.079 $\mu$$_B$ induced ferromagnetism along {\bf -b} with the {\it a} components again cancelling. But for the {\bf a} orientation the {\it b} and {\it c} components both cancel and the crystal remains antiferromagnetic. The fact that in no case is there any ferromagnetism along {\it a} complicates the interpretation of the 0.012 $\mu$$_B$ measured value \cite{Kim2006} which was obtained by applying a magnetic field along one of the in-plane $<$100$>$ directions of the substrate. It is likely, that to reduce the strain, there are domains in which {\it a} is in the $<$100$>$ direction and others in which it is in the $<$001$>$ direction which would make the experimental film value consistent with the theoretical value averaging over different domains though the theoretical value is a little bigger.  Additionally, we expect reduced energy denominators, caused by GGA density functional  gap narrowing, to enhance the SOC and therefore enhance the calculated SOC induced ferromagnetism, thus making the disagreement between the calculated and experimental ferromagnetism understandable. In fact the GGA+{\it U} energy gap is about twice \cite{23} that of the GGA, resulting in initially {\bf b}({\bf c}) antiferromagnetic spins having 0.049 (0.053) $\mu$$_B$ induced ferromagnetism in the {\bf c}({\bf b}) direction.

\begin{table}
\begin{ruledtabular}
\caption{Local spin and orbital moment components for different initial antiferromagnetic spin orientations in the GGA.}
\begin{tabular}{cccccccc}
orientation&  &         &  spin   &         & orbit & moment&       \\
\hline
      &       &    $a$    &    $b$    &    $c$    &   $a$   &   $b$  &$c$       \\
\hline
{\bf b}     & Cr1   &  0.013  & -2.637  & -0.073  &-0.001 & 0.042&0.007   \\
            & Cr2   & -0.013  &  2.637  & -0.073  & 0.001 &-0.042&0.007   \\
            & Cr3   &  0.013  &  2.637  & -0.073  &-0.001 &-0.042&0.007   \\
            & Cr4   & -0.013  & -2.637  & -0.073  & 0.001 & 0.042&0.007   \\
            & Total & 0       & 0       & -0.292  &0      & 0    &0.028   \\
{\bf a}     & Cr1   & 2.637   & 0.028   &-0.058   &-0.044 &-0.004&-0.005  \\
            & Cr2   &-2.637   &-0.028   &-0.058   & 0.044 & 0.004&-0.005  \\
            & Cr3   &-2.637   & 0.028   & 0.058   & 0.044 &-0.004&-0.005  \\
            & Cr4   & 2.637   &-0.028   & 0.058   &-0.044 & 0.004&-0.005  \\
            & Total & 0       & 0       & 0       &  0    & 0    & 0      \\
{\bf c}     & Cr1   & 0.043   & -0.079  & 2.637   &  0    & 0.009&-0.040  \\    
            & Cr2   & 0.043   & -0.079  &-2.637   &  0    & 0.009& 0.040  \\
            & Cr3   &-0.043   & -0.079  &-2.637   &  0    & 0.009& 0.040  \\
            & Cr4   &-0.043   & -0.079  & 2.637   &  0    & 0.009&-0.040  \\
            & Total & 0       & -0.316  & 0       &  0    & 0.036& 0      \\ 
\end{tabular}
\end{ruledtabular}
\end{table}   

    We are now in a position to see how the D-M theory applies to the current case. We note that the {\it a} coordinate is unique in that no weak ferromagnetism is induced along it if  the antiferromagnetic spins oriented along either of the other axes, or when {\it a} is the antiferromagnetic axis, there is no weak ferromagnetism along the other two axes. This uniqueness is a consequence of the different non primitive translations \cite{24} associated with the two-fold rotations about the different axes. The translations belonging to the {\it b} and {\it c} rotations interchange (G type) spin sublattices whereas the translation belonging to the {\it a} rotation does not. The unitary {\it Pnma} symmetry is preserved only when the antiferromagnetic polarization is oriented along {\it a}. Because each component {\it M$_{\alpha}$} of the ferromagnetic vector {\bf M} is rotated into {\it -M$_{\alpha}$} by one of the perpendicular two-fold rotations (the nonprimitive translations have no effect on {\bf M}), ferromagnetism is forbidden (unless there is a broken symmetry). When the antiferromagnetic polarization is oriented along the {\it b} axis, the unitary group is reduced to $E$, $I$, $\sigma_c$[$a$/2,0,$c$/2], and $U_c$[$a$/2,0,$c$/2]. These  four operations plus $\tau$, the time reversal operator, multiplying the remaining four reflection and rotation operators, $\sigma_b$[0,$b$/2,0], $U_b$[0,$b$/2,0], $\sigma_a$[$a$/2,$b$/2,$c$/2], and $U_a$[$a$/2,$b$/2,$c$/2] form the antiferromagnetic antiunitary group of the crystal. All eight operations leave the {\it c} (but not {\it a}) component of the spins' sign unchanged. If the antiferromagnetic polarization is along {\it c}, interchange the {\it b} and {\it c} subscripts to obtain the magnetic group. Then all eight operations leave the {\it b} (but not {\it a}) spin component unchanged. Therefore, all operations of the magnetic group are consistent with weak ferromagnetism along the {\it b} or {\it c} axis, respectively, if the antiferromagnetic spin orientation is along the {\it c} or {\it b} axis. One can also show that the antiferromagnetic spin tilts belong to the magnetic group \cite{25}. If a D-M
{\bf D}$\cdot${\bf S$_1$}$\times${\bf S$_2$} term is present in the Hamiltonian, it is easy to see that {\bf D} must be polarized in the {\bf a} direction, irrespective of the orientation of the spins.

    In summary, we have performed density functional calculations for a BCO film with lattice constants determined by the STO substrate on which it was epitaxially grown as well as for a bulk crystal with relaxed lattice constants, and found in both cases that they have the {\it Pnma} space group of the high temperature phase, indicating that the high temperature phase is stable (when grown epitaxially on STO) down to the 10 K to which the film was taken. We obtained a value for the mixed A and G type antiferroelectric Bi displacements; We found G type antiferromagnetism and if the antiferromagnetic spins were aligned along the $b(c)$ axis a weak ferromagnetic moment was spin-orbit induced along the $c(b)$ axis and if they were aligned along the $a$ axis there was no induced ferromagnetism. The uniqueness of the {\it a} axis was attributed to the nonprimitive translation accompanying its 2-fold rotation which does not interchange the spin sublattices. It was found that SOC induced antiferromagnetic spin tilting, consistent with the magnetic space group, always occurred where ferromagnetic tilting did not.  We showed how this could have been predicted from a theoretical analysis of the {\it Pnma} group, and that {\bf D} in the D-M Hamiltonian must point along the {\bf a} axis irrespective of the antiferromagnetic spin polarization, Note, however, that does not account for the antiferromagnetic spin tilting.

\begin{acknowledgments}
This work was supported by the Texas Advanced Computing Center TACC) and by The Welch Foundation under Grant No. F-0934. Y.Y also thanks to NSFC (10674163, 10974231) and the MOST Project (2007CB925000).
\end{acknowledgments}


\begin{thebibliography}{99}
\bibitem{23} The density functional error in the energy gap is usually about the same for GGA and GGA+$U$. The difference here arises from the isolated band of d states which form the top of the GGA DOS. and which are pulled into the continuum below by $U$.
\end{thebibliography}
\end{document}